# Universal Scaling Formalism and Analytical Optimization Criterion for Multiscale Geometric Design of Thermoelectric Metamaterials

Xanthippi Zianni

Dept. of Aerospace Science and Technology, National and Kapodistrian University of Athens, Greece.

E-mail: xzianni@aerospace.uoa.gr

## Abstract

Thermoelectric (TE) generators can directly convert heat into electricity, but their performance is often constrained by limited temperature gradients. Here it is shown that width-modulated metamaterials with constrictions and expansions (*constricted* geometries) enhance temperature difference $\Delta T$ by reduced *Transmissivity* ($Tr$), a geometry-based parameter defined by the ratio of constriction to expansion cross-sections. A *universal scaling behavior* of transport and key TE efficiency metrics with *Transmissivity* is demonstrated, spanning from the nanoscale to the macroscale. Analytical formalism validated through finite element calculations for a range of modulation geometries reveals that $\Delta T$, electrical and thermal resistances, efficiency, and power output are governed by a single scaling function, $g(Tr)$, independent of carrier type, material, or operating conditions. This function represents the conductance of a *constricted* geometry relative to a uniform-width counterpart. The developed framework yields *TE Performance Design Maps* and an analytical criterion for optimal TE performance, with the maximum power density achieved at an optimal *Transmissivity $Tr_{opt}$*, determined by the condition that the functional $g(Tr_{opt})$ equals the Biot number, the dimensionless ratio $hL/k$ of the convection coefficient $h$, the structure length $L$ and the material thermal conductivity $k$. *Transmissivity* is established as a robust, multiscale design parameter - analogous to nature's hierarchical structures for optimized functionality. This work provides the theoretical framework for multiscale design and optimization of *constricted* geometries, thereby enabling systematic exploration of design strategies for next-generation TE modules based on advanced thermoelectric metamaterials.

## Keywords

metamaterials, geometry modulation, universal scaling, transmissivity, thermoelectric performance, thermal management, thermoelectric modules

# 1. Introduction

Thermoelectric (TE) energy conversion directly transforms heat into electricity, offering a solid-state solution to energy and environmental challenges[1-3]. Thermoelectric devices, with their compact and scalable design, offer a sustainable response to the growing power demands of microelectronics, AI, autonomous sensors, and next-generation technologies by directly converting heat into electricity and recycling waste heat[3-10].

The performance of TE devices is governed by a combination of properties of the material constituting the TE leg - primarily the figure of merit ($ZT$), which depends on the Seebeck coefficient, the electrical conductivity, and the thermal conductivity - as well as device-level factors such as maintenance of high temperature gradients with the use of efficient heat exchangers[2,11,12], low contact/interface resistances to decrease output power losses[1,2,13-16] and alternatives to the traditional π-module geometry to meet the needs of various applications [12,17-19]. Although materials with high $ZT$ values have been discovered, their integration into efficient modules remains limited due to unresolved technological challenges[16]. Today, a revolutionary shift in TE research recognizes material geometry as an independent factor capable of maximizing the thermal resistance of the TE leg and maintaining a high temperature difference, $\Delta T$, under various operating conditions.

Optimizing conventional TE leg geometry involves adjusting the height and cross-sectional area of a cuboid-shaped material to balance the requirements for high thermal resistance and low electrical resistance[2,14,20]. Non-cuboid shapes with variable cross-section - particularly *pyramidal* geometries - have drawn research interest in TEs since the early years of research[21-23], but remained of limited focus until recent advances in metamaterials science and technology renewed attention. Width-modulated metamaterials featuring constrictions and expansions (*constricted* geometries) were proposed to increase the TE efficiency by geometrical tuning the electrical and thermal transport[24-28]. Research interest in *pyramidal*[29-32] and *constricted* geometries[33-35] has been reignited, theoretically demonstrating enhanced TE performance. Furthermore, significant progress in advanced manufacturing techniques - such as additive manufacturing and Direct Ink Writing - enabled fabrication of non-conventional TE legs and initial experimental validation of predictions[36-37]. These findings sparked a surge of research on non-cuboid TE leg configurations[38-53]. Future advances will depend on multi-factor optimization of TE devices based on non-cuboid legs[16,54,55]. Optimizing metamaterial geometry with respect to TE performance metrics is a central prerequisite for advancing thermoelectric devices. Numerical studies have examined a wide range of non-cuboid shapes, demonstrating that variable cross-sections can improve performance under diverse boundary conditions and constraints (e.g., constant volume, constant surface area)[33,35,37,43,46,56–58]. Among these, constricted geometries consistently outperform cuboid and other variable cross-section designs[34,35,37,49]. Recent proposals for novel TE leg architectures - enabled by advanced manufacturing techniques - underscore the growing emphasis on enhancing TEG efficiency through geometry optimization, informed by high-fidelity simulations[16].

Despite this progress, three key challenges persist[54]: (i) an incomplete understanding of how shape governs the thermal and electrical resistances of metamaterials, (ii) the lack of validated optimization criteria applicable across different geometries, and (iii) the absence of standardized descriptors required for AI-driven design. The present work addresses these challenges introducing a physically grounded analysis supported by analytical formalism validated through finite element calculations across a range of modulation profiles, which yields an optimization criterion for maximizing TE performance in *constricted* metamaterial geometries.

Geometry modulation affects transport on two distinct levels: material and structural. At the material level, it influences electric and thermal transport by modifying the energy states of electrons and phonons and altering their scattering[24–26,59]. At the structural level, transport is constrained by *reduced Transmissivity* - a mechanism arising purely from geometry, and fundamentally different from conventional boundary or interface scattering mechanisms[27].

The concept of *Transmissivity*, fully defined by the geometry-modulation profile, was first introduced as an intuitive framework for analyzing nanoscale thermal transport in *constricted* metamaterials[27]. Follow-up studies employing phonon Monte Carlo simulations validated these initial insights and uncovered distinct features associated with this mechanism[28]. A central finding was the characteristic scaling between thermal conduction and *Transmissivity*[27,28]. In this work, this concept is extended by demonstrating a *universal scaling behavior* of electrical transport, thermal transport, and TE performance metrics with respect to *Transmissivity*, spanning from the nanoscale to the macroscale. This establishes *Transmissivity* as a fundamental geometric descriptor—one that enables general design rules and global optimization criteria for enhancing TE performance.

Section 2 introduces the theoretical model and methodology. Section 3 presents and discusses the simulation results along with the analytical formalism, establishing universal scaling relations for electrical resistance, thermal resistance, temperature difference, efficiency and maximum output power with *Transmissivity*. The section concludes by deriving *Performance Design Maps* and formulating a global optimization criterion for maximizing the TE performance of *metamaterials* with *constricted* geometry.

## 2. Method and validation

Calculations were carried out using the finite element method where the structure is discretized into individual cells. At two opposite sides of the material are imposed electrical potential difference $\Delta V$ and temperature difference $\Delta T$. The electrical current densities, $\boldsymbol{J}$, and heat flux, $\boldsymbol{q}$, are calculated using the following definitions[2,14]:

$$\boldsymbol{J} = -\sigma(\nabla V + S\,\nabla T) \tag{1}$$

$$\boldsymbol{q} = -k\,\nabla T + ST\boldsymbol{J} \tag{2}$$

where $\sigma$ is the electrical conductivity, $S$ is the Seebeck coefficient and $k$ is the thermal conductivity of the material. The electrical potential, $V$, and the absolute temperature, $T$, profiles are determined applying electric current continuity and energy conservation conditions:

$$\nabla \cdot \boldsymbol{J} = 0 \tag{3}$$

$$\nabla \cdot \boldsymbol{q} = -\nabla V \cdot \boldsymbol{J} \tag{4}$$

Calculations performed across different materials, dimensions, and temperature gradients consistently revealed the same trends. For clarity, we present representative results for n-type $Bi_2Te_{2.7}Se_{0.3}$ *constricted* geometries with single- and multiple- modulation profiles, fully accounting for the experimentally measured[60] temperature dependence of its electrical conductivity, Seebeck coefficient, and thermal conductivity. The hot-side contact temperature is fixed at $T_h$=400 K, while the ambient temperature is set at $T_a$=300 K. Simulation results are interpreted by analytical formalism; their mutual comparison provides validation for both approaches.

# 3. Results and Discussion

### 3.1. Scaling of transport with *Transmissivity.*

Evaluation of a TE device performance typically relies on the figure of merit *ZT*, the conversion efficiency, *η,* and the maximum output power, $P_{max}$, of the material constituting the TE leg[2]:

$$ZT = \frac{\sigma S^2 T}{k} \tag{5}$$

$$\eta = \frac{P_{el}}{Q_{in}} \tag{6}$$

$$\eta_{max} = \frac{\Delta T}{T_h} \frac{\sqrt{1+ZT}-1}{\sqrt{1+ZT}+\frac{T_c}{T_h}} \tag{6a}$$

$$P_{max} = \frac{V_{OC}^2}{4\,R_{el}} = \frac{S^2 \Delta T^2}{4\,R_{el}} \tag{7}$$

where *σ, S, k* are respectively the material electrical conductivity, Seebeck coefficient and thermal conductivity. $P_{el}$ is the electrical output power and $Q_{in}$ is the incoming thermal power. $V_{OC}$ is the open circuit voltage, $T_{h(c)}$ is the temperature of the hot (cold) side of the material, *ΔT* is the temperature difference across the structure and $R_{el}$ is the material electrical resistance.

To improve TE efficiency, it is essential to enhance the intrinsic material's figure of merit *ZT* and maintain a large temperature difference *ΔT* across the material [Eq. (6a)]. The value of *ZT* can be increased by engineering the material microscale morphology to optimize the transport properties: *σ, S*, and *k*. Preserving a high *ΔT* requires minimizing the thermal conductance of the material $G_{th}$ (or equivalently, maximizing the material's thermal resistance $R_{th}$), thereby reducing parasitic heat flow. This reduction can be achieved by lowering the material's intrinsic thermal conductivity at the microscale or by geometrically tuning $G_{th}$ ($R_{th}$). Among various geometry design strategies, *constricted* geometry is a suitable approach because width-modulated materials by constrictions exhibit lower $G_{th}$ than the corresponding $G_{th}^0$ of constant-width cuboids. This was initially demonstrated in width-modulated nanostructures[27,28]. In particular, prior studies demonstrated that the thermal conductance $G_{th}$ of width-modulated nanoslabs (wires and films) decreases monotonically as the constriction width is reduced, relative to the conductance $G_{th}^0$ of the corresponding uniform structure with constant width. For multiple-constriction modulation profiles (Fig. 1a), this monotonic reduction follows a simple analytical relation[28]:

$$\frac{G_{th}}{G_{th}^0} \approx Tr \qquad (8)$$

$$Tr = \frac{A_C}{A} \qquad (9)$$

where $A$ and $A_C$ are the cross-sectional areas of the expansions and the constrictions respectively (Fig. 1). $Tr$ defined as the ratio of the two characteristic cross-sections of the modulated material, expresses the actual *constricted* geometry *Transmissivity*[27,28]. This definition should be appropriately modified for other metamaterial geometries.

Eq. (8) makes it explicit that in *constricted* geometries $G_{th}$ is smaller than $G_{th}^0$ because the constriction area $A_C$ is smaller than the expansion area $A$ ($Tr < 1$). Moreover, this equation shows that the ratio $G_{th}/G_{th}^0$ scales directly with $Tr$, i.e. the decrease in $G_{th}$ relative to $G_{th}^0$ is governed by the ratio $A_C/A$. Consequently, the relative decrease in thermal conductance induced by geometric modulation will be identical across structures with different absolute values of $A$ (or $A_C$) as long as they have the same $Tr$.

The scaling of thermal conductance with *Transmissivity* holds irrespective of the modulation profile, although the exact functional dependence is dictated by the specific form of that profile[28,61,62]. Eq. (8) can therefore be recast in a more general form to account for this dependence:

$$\frac{G_{th}}{G_{th}^0} = g(Tr) \qquad (10)$$

where $g(Tr)$ denotes the functional dependence of $G_{th}/G_{th}^0$ on $Tr$.

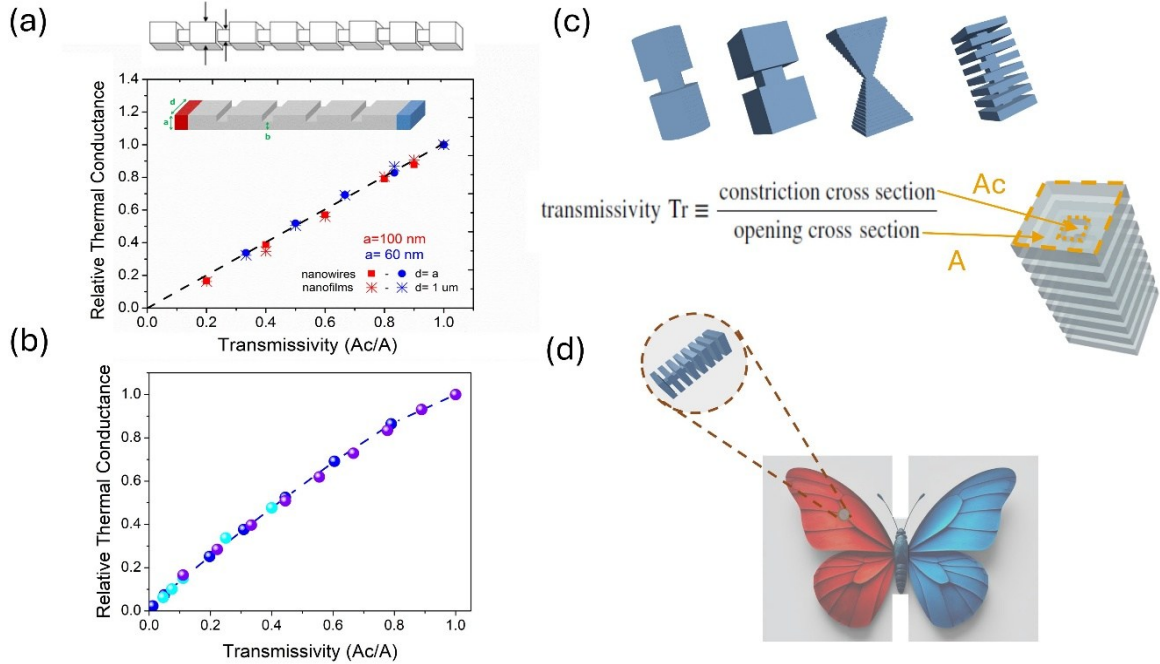

**Fig. 1. Multiscale Scaling with Transmissivity**. Thermal conductance of width-modulated metamaterials versus *Transmissivity* (a) at the nanoscale[28], and (b) at the macroscale, for dimensions detailed in Fig. 2. Thermal conductance is expressed relative to that of the constant-width material. (c) Definition of *Transmissivity* as the ratio of the cross-sectional area of the constriction $A_C$ to that of the expansion $A$. (d) Artistic representation of the concept of multiscale design of *constricted* metamaterials for energy convertors, reminiscent of nature-inspired designs that employ hierarchical scaling for optimized functionality.

The reduction of $G_{th}$ in *constricted* metamaterials, relative to $G_{th}^0$, can be fully attributed to the decreased *Transmissivity* imposed by the modulation geometry. The scaling dependence of the relative conductance $G_{th}/G_{th}^0$ on $Tr$ – a quantity determined solely by geometry - reflects the geometric origin of the reduced thermal conductance in this class of metamaterials. This scaling behavior has been consistently observed across different nanoscale modulation profiles[28,61,62].

Here, this result is extended to the macroscale, demonstrating the same behavior in *constricted* materials with millimetre-scale dimensions, as typically found in TE legs. Finite element calculations were performed varying cross-sectional areas $A$ and $A_C$. Representative results are shown for single-constriction (Fig. 2) and multiple-constriction modulation profiles (Fig 3), under both one- (1D) and two- (2D) dimensional width-modulation schemes: (i) varying $A$ while keeping $A_C$ fixed, and (ii) varying $A_C$ with $A$ held constant. In all cases, the simulations confirm that $G_{th}/G_{th}^0$ scales universally with *Tr*.

Notably, electrical conductance exhibits the same dependence on *Tr* as thermal conductance. Regardless of the absolute values of thermal or electrical resistance, the normalized ratio $G_{th(el)}/G_{th(el)}^0$ follow the same functional form *g(Tr)* (Fig. 4). This universality observed across all modulation profiles (Figs. 2-4) indicates that the impact of geometry modulation on transport is inherently geometric and independent of the nature of carriers. Hence, it holds:

$$\frac{G_{el}}{G_{el}^0} = g(Tr) \tag{11}$$

Eqs. (10) and (11) make explicit that, in *constricted* metamaterials as in uniform materials, electrical and thermal conduction remain coupled, meaning that geometry modulation does not offer a way around the persistent trade-off between them.

The functional form *g(Tr)* is determined by the specific modulation profile. Both nanoscale[28,61,62] and macroscale calculations, indicate the following approximate relation:

$$g(Tr) \sim Tr^n \tag{12}$$

with $n \approx 1$ for multiple-constriction modulation and $n \approx 0.5$ for single-constriciton modulation.

The relative conductance $G_{th(el)}/G_{th(el)}^0$ decreases monotonically with decreasing *Tr* (Figs. 2b,3b). In contrast, the absolute values of conductance (or resistance) may not vary monotonically with *Tr*, as the outcome depends on which dimension is held constant. This effect is illustrated in Fig.4 where resistance is shown in both absolute and relative terms. Specifically, resistance increases with decreasing *Tr* when $A_c$ decreases at fixed *A*, whereas it decreases with decreasing *Tr* when *A* increases at fixed *Ac*. These opposite trends arise because, in the first case, $R_{th(el)}^0$ remains constant, and resistance is governed solely by the variation in *Tr* due to varying $A_C$. In the second case, however, $R_{th(el)}^0$ itself varies with *A*, so resistance is primarily determined by the increase in $R_{th(el)}^0$ as *A* decreases.

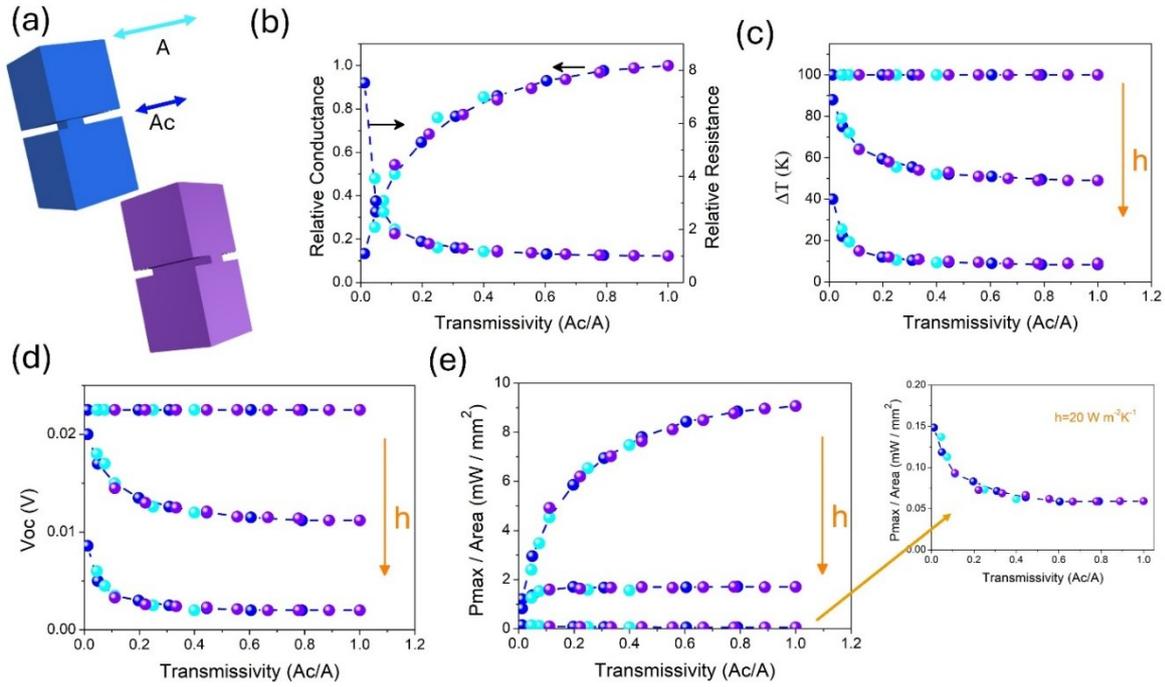

**Fig. 2. Scaling behavior for single-constriction width-modulation.** (a) Calculations for structures with 1D (purple) and 2D (blue and cyan) width-modulation geometries. (b) Relative electrical and thermal conductance/resistance. (c) Temperature difference $\Delta T$. (d) Open-circuit voltage $V_{OC}$. (e) Maximum output power density $P_{max}/A$, under three operating conditions with decreasing convection coefficient: $h = \infty$ (fixed $\Delta T$), 200 Wm$^{-2}$K$^{-1}$, and 20 Wm$^{-2}$K$^{-1}$. The inset in (e) corresponds to $h = 20$ Wm$^{-2}$K$^{-1}$. Two modulation schemes are compared: fixed $A = 4$ mm×4 mm with variable $Ac$ (blue and purple symbols), and fixed $Ac = 2$ mm×2 mm with variable $A$ (cyan symbols). The structure length $L$ is fixed at 8 mm.

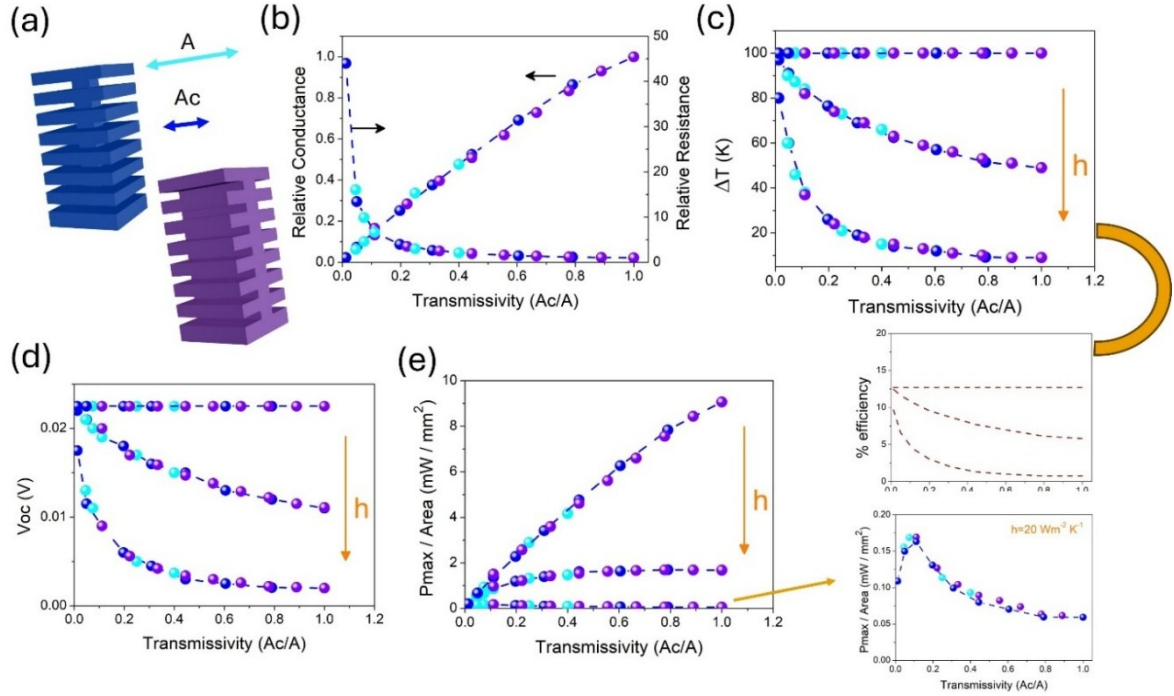

**Fig. 3. Scaling behavior for multiple-constriction width-modulation.** (a) Calculations for structures with 1D (purple) and 2D (blue and cyan) width-modulation geometries. (b) Relative electrical and thermal conductance/resistance. (c) Temperature difference $\Delta T$ and efficiency $n_{max}$. (d) Open-circuit voltage $V_{OC}$. (e) Maximum output power density $P_{max}/A$, under three operating conditions with decreasing convection coefficient: $h = \infty$ (fixed $\Delta T$), 200 Wm$^{-2}$K$^{-1}$, and 20 Wm$^{-2}$K$^{-1}$. The inset in (e) corresponds to $h= 20$ Wm$^{-2}$K$^{-1}$. Colores in symbols correspond to the two modulation schemes described in the Fig.2 caption, using the same structural dimensions as in Fig. 2.

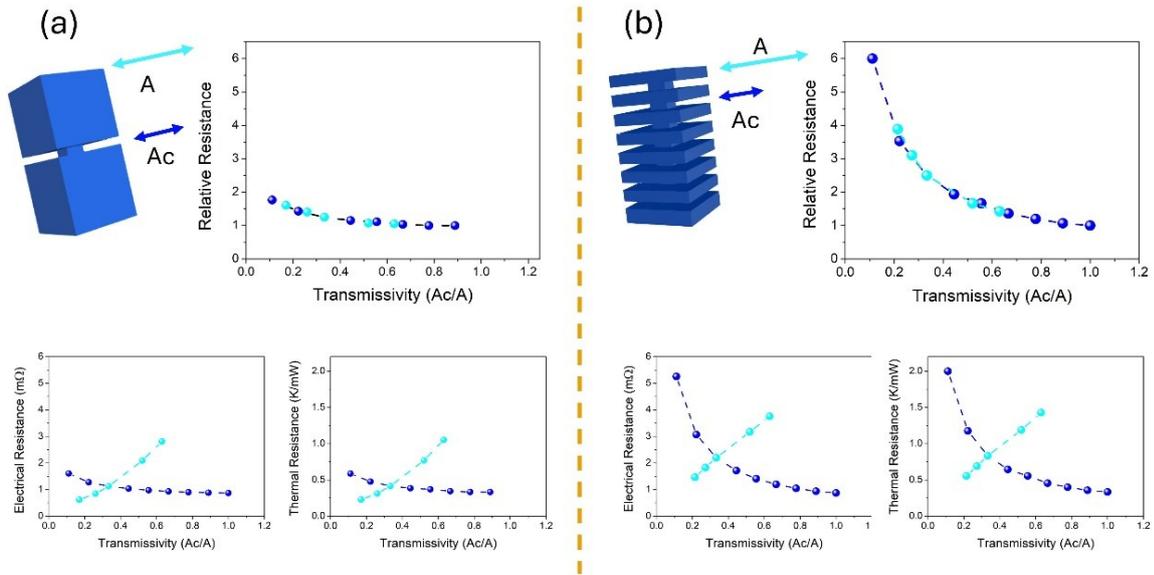

**Fig. 4. Scaling of resistance with Transmissivity.** Electrical and thermal resistances – shown both in absolute terms and normalized to the constant cross-section structure ($A$) ($R_{th(el)}/R_{th(el)}^0$) – as a function of *Transmissivity* for width-modulated structures with: (a) single-constriction and (b) multiple-constriction profiles. Blue and cyan symbols correspond to the two modulation schemes described in the Fig.2 caption, using the same structural dimensions as in Fig. 2.

The demonstrated scaling dependence has been validated across different materials and temperature gradients, confirming that reduced *Transmissivity* acts as a purely geometric mechanism for tuning conduction. This mechanism is thus distinct from conventional mechanisms such as intrinsic scattering or boundary/interface roughness scattering, which depend on carrier type and intrinsic material properties. The consistent scaling behavior of transport with *Transmissivity*—across length scales from the nanoscale to the macroscale—demonstrates that *Transmissivity* reflects a geometric property of the *constricted* metamaterial rather than a simple geometric ratio. For example, assigning the same numerical ratio $A_C/A$ to *pyramidal* geometries would not capture the physical essence of *Transmissivity*. Owing to this scaling relationship, the reduction (increase) in conductance (resistance) in a *constricted* metamaterial relative to the corresponding uniform material can be fully determined by this geometric property, positioning *Transmissivity* as a robust descriptor of the effect of geometry modulation on transport across multiple scales.

These results establish that conduction in *constricted* metamaterials exhibits a *universal scaling dependence* on *Transmissivity* – independent of transport carrier (electrons or phonons), intrinsic material properties, or specific modulation profile.

## 3.2 Scaling of *ΔT* and TE efficiency with *Transmissivity*.

The efficiency of a TE module is governed by the temperature difference *ΔT* across the leg material [Eq. (6a)]. This difference reaches its maximum under fixed contact temperature conditions. In practical operating environments, however, contact temperatures often fluctuate due to convective heat exchange. Under such convective conditions, the contact temperature is determined by both the thermal resistance $R_{th}$ of the leg and the strength of convective flow - quantified by the convection heat transfer coefficient, *h*. For uniform materials, it can be shown (Appendix A) that:

$$\Delta T = \frac{hL}{k + hL} \Delta T_{max} \tag{13}$$

where $\Delta T_{max} \equiv T_h - T_a$, $T_a$ being the temperature of the ambient. *L* denotes the length of the material along the temperature gradient.

A higher value of *h* corresponds to stronger heat exchange, causing the contact temperature to approach the ambient. In the theoretical limit *h*→∞, the contact temperature equals that of the environment yielding $\Delta T = \Delta T_{max}$. As *h* decreases, the attainable *ΔT* is reduced, a well-known challenge for TE devices operating under weak convection. This relationship is captured by Eq. (13) and is confirmed by finite element calculations for constant-width geometries (*Ac=A* and *Tr*=1) (Figs. 2c and 3c). The thermal conductivity is weakly temperature dependent[60]. Simulations show exact quantitative agreement with Eq. (13) when using the average thermal conductivity *k*= 1.8 Wm$^{-1}$K$^{-1}$ and the length *L*=8 mm of the simulated structures.

Finite element calculations indicate that, under convective conditions *ΔT* is higher in *constricted* materials (*Ac<A* and *Tr*<1) than in the corresponding uniform material with constant cross-section *A* (Figs. 2c and 3c). This demonstrates that the temperature difference *ΔT* diminishes less (is preserved better) in *constricted* geometries compared to their cuboid counterparts under the same convective operating conditions. Importantly, *ΔT* increases monotonically as *Tr* decreases, reflecting the concurrent increase of the relative thermal resistance $(R_{th}/R_{th}^0)$ and exhibits a direct scaling relationship with *Tr*. These simulation results are further validated by the following analytical formalism obtained by extending Eq. (13) to non-uniform materials (Appendix A):

$$\Delta T = \frac{hL/k}{G_{th}/G_{th}^0 + hL/k} \Delta T_{max} \qquad (14)$$

Eq. (14) and finite element calculations show quantitative agreement across all simulated structures ($Tr \leq 1$). The simulation results in Figs. 2c and 3c match exactly the predictions of Eq. (14) when using $k$= 1.8 Wm$^{-1}$K$^{-1}$, $L$=8 mm and the calculated values of $G_{th}/G_{th}^0$ in Figs. 2b and 3b.

Eq. (14) makes clear that the enhancement of *ΔT* arises from the reduced relative thermal conductance $G_{th}/G_{th}^0$, which is directly governed by the lower *Transmissivity* of the *constricted* geometry. It also interprets the scaling of *ΔT* with *Tr* observed in the simulations, showing that it originates from the *scaling* dependence of $G_{th}/G_{th}^0$ on *Tr*. A key implication of this scaling is that structures with identical modulation ratios (*Ac/A*) - and thus the same *Tr* – will exhibit the same *ΔT*, regardless of their absolute cross-sectional areas. This leads to the important conclusion that, for fixed *h*, structure length *L* and intrinsic material *k*, *ΔT* is determined by the *Transmissivity* of the *constricted* geometry.

A comparison of the results for a given *h* (Figs. 2c and 3c) shows that *ΔT* rises more sharply with decreasing *Tr* in multiple-constriction modulations than in single-constriction cases. This behavior stems from the fact that $G_{th}/G_{th}^0$ is lower in structures with multiple constrictions than in those with a single constriction at the same *Tr* (Figs. 2b and 3b). As the number of constrictions increases, conductance decreases (and resistance increases) progressively with the degree of modulation[62]. Consequently, geometries with more extended modulation yield a more pronounced increase in *ΔT* as *Tr* decreases.

At the macroscale, the figure of merit *ZT* remains constant since the intrinsic material transport properties ($\sigma$, *k* and *S*) are independent of the material dimensions. As a result, the TE efficiency directly follows the behavior of *ΔT* [Eq. (6a)]. The maximum efficiency *n$_{max}$* increases monotonically as *Tr* increases (Fig. 3c). Like *ΔT*, *n$_{max}$* also scales with *Tr*, owing to its dependence on the ratio $G_{th}/G_{th}^0$, as captured by the following analytical expression derived from Eqs. (6a) and (14):

$$\eta_{max} = \frac{\Delta T_{max}}{T_h} \frac{\sqrt{1+ZT}-1}{\sqrt{1+ZT}+1-\frac{\Delta T_{max}}{T_h}\frac{hL/k}{G_{th}/G_{th}^0+hL/k}} \frac{hL/k}{G_{th}/G_{th}^0 + hL/k} \qquad (15)$$

From Eqs. (10),(14) and (15), universal scaling relations for *ΔT* and *n$_{max}$* can be expressed in terms of the same function *g(Tr)*:

$$\Delta T\ (Tr) = \frac{hL/k}{g(Tr) + hL/k} \Delta T_{max} \qquad (16)$$

$$\eta_{max}(Tr) = \frac{\Delta T_{max}}{T_h} \frac{\sqrt{1+ZT}-1}{\sqrt{1+ZT}+1-\frac{\Delta T_{max}}{T_h}\frac{hL/k}{g(Tr)+hL/k}} \frac{hL/k}{g(Tr)+hL/k} \qquad (17)$$

The optimization of the *constricted* geometry is strongly dictated by its *Transmissivity*, which serves as a key design parameter. As indicated by Eqs. (6a) and (14) enhancement in TE efficiency arises primarily from the elevated thermal resistance $R_{th}$, compared with $R_{th}^0$. As indicated by Eq. (14), the increase (decrease) in $R_{th}$ ($G_{th}$) directly amplifies the temperature difference $\Delta T$ under convective operation. To explore the role of the constriction, the data of Fig.2c for $\Delta T$ due to a single-constriction modulation are re-plotted against the *Inverse Transmissivity* ($Tr^{-1}$) (Fig. 5b). Then, a distinct progression is observed regardless of the specific constriction profile - whether abrupt or gradual (Fig. 5a): an abrupt growth in $\Delta T$ is followed by saturation, where the plateau asymptotically approaches the ambient temperature. This saturation limit is prescribed by Eq. (14), which identifies the convection coefficient $h$, material conductivity $k$, and length $L$ as the key determinants of the plateau height. The emergence and extent of the plateau are governed by the functional $g(Tr)$ [Eq.(16)], thereby linking the phenomenon directly to geometric *Transmissivity*.

Analysis of the temperature distribution (T-profile) across the *constricted* material provides further insight into the underlying mechanism for the occurrence of the plateau (Fig. 5b). For large $Tr$, the T-profile is nearly linear. As $Tr$ decreases, however, it progressively distorts and eventually develops into a broad, stable window centered at the constriction. By Fourier's law, thermal conductivity is inversely related to the local temperature gradient. Thus, the sharp gradient appearing at the constriction at the onset of the plateau signals the formation of a dominant thermal resistance. This localized resistance, referred to as *Constriction Thermal Resistance* (*CTR*), first identified at the nanoscale (Fig.5d)[28] and demonstrated here at the macroscale (Fig. 5c) - remains nearly constant within the plateau regime, as indicated by the weak variation in temperature profiles across different $Tr$ values. The formation of *CTR* accounts for the abrupt rise in $\Delta T$ below a critical $Tr$. As a result, maximum TE efficiency is achieved at the onset of this plateau, marking the transition into the *CTR*-dominated regime.

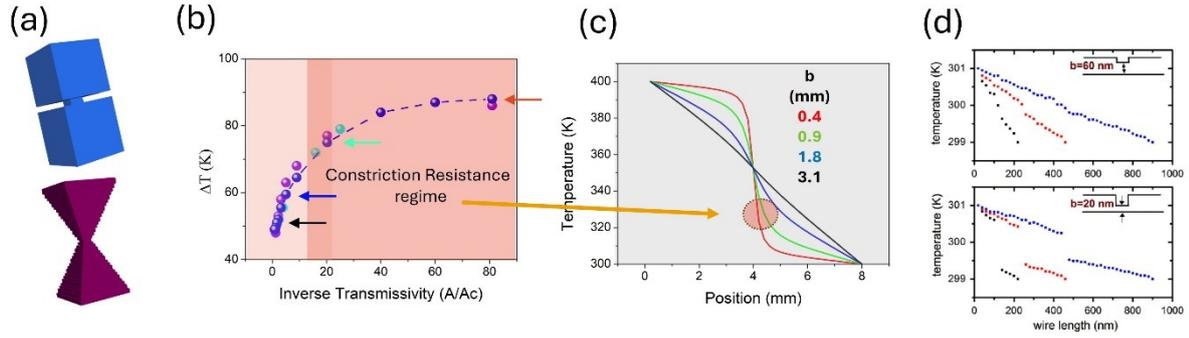

**Figure 5. *Constriction Thermal Resistance* across scales.** (a) Abrupt (blue) and smooth (violet) constriction geometries. (b) Temperature difference $\Delta T$ vs. *Inverse Transmissivity* ($Tr^{-1}$) for single-constriction structures: fixed $A$ with variable $Ac$ (blue, purple and violet), and fixed $Ac$ and variable $A$ (cyan), using the same dimensions as in Fig.2. (c) Formation of *Constriction Thermal Resistance (CTR)* at the macroscale. (d) Corresponding CTR formation at the nanoscale (adapted from Ref.28).

### 3.2 Scaling of the output power density with *Transmissivity*.

The maximum output power $P_{max}$ is governed by the trade-off between the open-circuit voltage $V_{OC}$ and the electrical resistance $R_{el}$ [Eq. (7)]. $V_{OC}$ is directly proportional to $\Delta T$ following the relation: $V_{OC}=S \cdot \Delta T$. The calculated values of $V_{OC}$ (Figs. 2d and 3d) are fully consistent with this relation. Moreover, the spatial distribution of voltage across the *constricted* material mirrors the corresponding temperature distribution (Fig. 6).

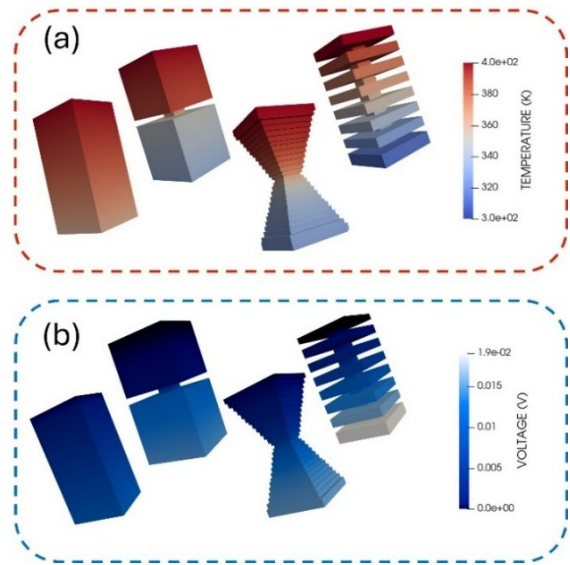

**Fig. 6. Temperature and voltage distribution.** (a) Temperature and (b) voltage profiles across different *constricted* geometries assuming identical material properties, structural dimensions and convective operating conditions.

Under fixed $\Delta T$, $P_{max}$ of a *constricted* material is lower than that of the corresponding uniform material $P_{max}^0$. This occurs because $V_{OC}$ remains unchanged, while $R_{el}$ is reduced relative to $R_{el}^0$. This relationship is made explicit by the following expression derived from Eq. (7):

$$P_{max} = \frac{P_{max}^0}{R_{el}/R_{el}^0} = P_{max}^0 \, (G_{el}/G_{el}^0) = P_{max}^0 \, g(Tr) \tag{18}$$

$$P_{max}^0 = \frac{S^2 \Delta T_{max}^2}{4 \, R_{el}^0} \tag{19}$$

Eq. (18) shows that the ratio $P_{max}/P_{max}^0$ is directly proportional to $G_{el}/G_{el}^0$, and therefore decreases as $Tr$ is reduced. Furthermore, $P_{max}/P_{max}^0$ follows the same scaling dependence on $Tr$, governed by the function $g(Tr)$.

From Eq. (19) it is obtained:

$$P_{max}^0 = A \, \frac{\sigma S^2 \Delta T_{max}^2}{4 \, L} \tag{20}$$

Eqs. (18) and (20), give:

$$P_{max}/A = \frac{\sigma S^2 \Delta T_{max}^2}{4 \, L} \, g(Tr) \tag{21}$$

Eq. (21) demonstrates that the output power density $P_{max}/A$ scales with the same functional dependence as the relative conductance $G_{el}/G_{el}^0$, consistent with the simulation results (Figs. 2e and 3e).

Both $P_{max}/P_{max}^0$ and $P_{max}/A$ decrease monotonically with $Tr$ according to $g(Tr)$. However, the absolute value of $P_{max}$ does not necessarily vary monotonically with $Tr$, depending on which structural dimension is held constant – analogous to the case of $R_{el}$. As shown in Fig. 7, $P_{max}$ decreases with decreasing $Tr$ when $Ac$ decreases at fixed $A$, whereas it increases when $A$ increases at fixed $Ac$. These trends arise directly from the corresponding dependencies of $R_{el}$ (Fig. 4) and their previously discussed interpretation.

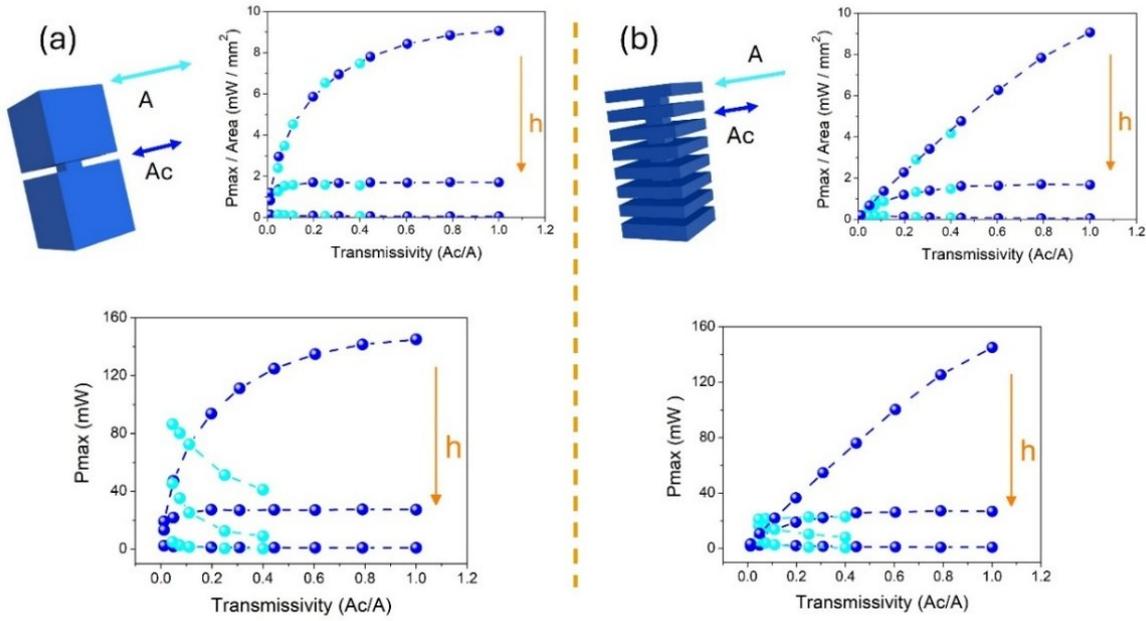

**Fig. 7. Scaling of output power with *Transmissivity*.** Absolute output power $P_{max}$ and output power density $P_{max}/A$ are shown for (a) single-constriction modulation, and (b) multiple-constriction modulation, as in Fig.4. Results are presented under three operating conditions are presented with decreasing convection coefficient $h$: fixed $\Delta T$ ($h=\infty$), $h = 200$ Wm$^{-2}$K$^{-1}$ and $h= 20$ Wm$^{-2}$K$^{-1}$.

Under convective operating conditions, the combination of Eqs. (7) and (14) yields:

$$P_{max} = P_{max}^0 \left[\frac{hL/k}{G_{th}/G_{th}^0 + hL/k}\right]^2 (G_{el}/G_{el}^0) \tag{22}$$

or equivalently, using Eqs. (10) and (11):

$$P_{max} = P_{max}^0 \left[\frac{hL/k}{g(Tr) + hL/k}\right]^2 g(Tr) \tag{23}$$

Then, the output power density $P_{max}/A$ can be expressed:

$$P_{max}/A = \frac{\sigma S^2 \Delta T_{max}^2}{4L} \left[\frac{hL/k}{G_{th}/G_{th}^0 + hL/k}\right]^2 (G_{el}/G_{el}^0) \tag{24}$$

or equivalently:

$$P_{max}/A = \frac{\sigma S^2 \Delta T_{max}^2}{4L} \left[\frac{hL/k}{g(Tr) + hL/k}\right]^2 g(Tr) \tag{25}$$

For uniform materials ($Tr=1$), $P_{max}$ decreases with decreasing convection coefficient $h$, due to the corresponding decrease of $\Delta T$ as expected by Eqs (7) and (14) and illustrated by finite element calculations in Fig. 7. $P_{max}/P_{max}^0$ and $P_{max}/A$ also decrease accordingly (Eqs. (22) and (24)). Moreover, they are functions of $g(Tr)$ and thus scale with *Transmissivity* (Eqs. (23) and (25)). Corresponding finite element calculations are shown in Figures 2e and 3e for $P_{max}/A$.

For constricted materials ($Tr<1$), $P_{max}$ may not vary monotonically with *Transmissivity* depending on which dimension is kept fixed similarly as in the case of $R_{el}$, similarly as for fixed $\Delta T$. Finite element calculations show that, contrary to the fixed $\Delta T$ case, under convective boundary conditions, $P_{max}/P_{max}^0$ and $P_{max}/A$ may show non-monotonic variation with $Tr$. Such a non-monotonic dependence is clearly shown in Fig. 3e for weak convection with $h=20$ Wm$^{-2}$K$^{-1}$. In this case $P_{max}/A$ shows a maximum at an optimal $Tr\sim0.09$. The non-monotonic dependence is depicted by the analytical formalism Eqs (23) and (25). These equations have global maximum when:

$$g(Tr_{opt}) = hL/k \tag{25}$$

at an optimal *Transmissivity* $Tr_{opt}$ where $P_{max}/P_{max}^o$ and $P_{max}/A$ reach their maxima. This result is confirmed by the finite element calculations. For example, applying this criterion to the dataset of Fig. 2b - for which $g(Tr) \approx Tr^{0.5}$ - yields $Tr_{opt}\sim0.8$ for $h=200$ Wm$^{-2}$K$^{-1}$ and $Tr_{opt}\sim0.008$ for $h=20$ Wm$^{-2}$K$^{-1}$, consistent with the simulation results (Fig.2e). Applying this criterion to the dataset of Fig. 3b - for which $g(Tr) \approx Tr$ - yields $Tr_{opt}\sim0.9$ for $h=200$ Wm$^{-2}$K$^{-1}$ and $Tr_{opt}\sim0.09$ for $h=20$ Wm$^{-2}$K$^{-1}$, consistent with the simulation results (Fig.3e).

Eq. (25) states that $P_{max}/P_{max}^0$ and $P_{max}/A$ are maximized when the functional $g(Tr)$ equals the Biot number $Bi$ ($=hL/k$). The Biot number quantifies the balance between external convection and internal conduction. For strong convection, $Bi \sim 1$ indicating that internal conduction is comparable to surface convection. In this regime, a significant temperature gradient is established across the material, and uniform-width geometry with high electrical conductance are optimal for maximizing TE output power. For weak convection, $Bi \ll 1$ indicating that surface convection dominates. In this regime, *constricted* geometries with low *Transmissivity* are preferred because they have decreased thermal conduction and thus preserve higher temperature gradient and enhanced output power.

The relevance of this condition for optimizing *constricted* geometries for maximum TE performance becomes even clearer when considering that: (i) *constricted* geometries consistently yield higher TE efficiency than uniform-width geometries due to enhanced *ΔT* under convective operating conditions, and (ii) the corresponding output power density does not necessarily increase, because while the reduced *Tr* boosts *ΔT*, it also increases the relative electrical resistance, partially offsetting the efficiency gains in efficiency. Therefore, maximizing TE performance ultimately requires maximizing output power density. In this context, Eq. (25) provides the optimization criterion for designing *constricted* geometries for maximum output power and maximum TE performance under convective operating conditions.

Eqs. (23), (25), and (26), along with their graphical representations (such as Figs. 2e and 3e) constitute analytical *TE Performance Desing Maps*. These maps provide practical, predictive tools for optimizing thermoelectric metamaterial geometries. They enable designers to determine, in advance, whether a certain *constricted* geometry will yield a net increase in output power, and if so, to quantify its magnitude for a given choice of geometric parameters and cross-sectional areas ($A$ and $A_C$).

## 4. Conclusions

A *universal scaling behavior* of transport and key TE performance metrics with *Transmissivity* is demonstrated in width-modulated metamaterials with constrictions and expansions, spanning from the nanoscale to the macroscale, using analytical formalism and simulations across a range of modulation profiles with single and multiple constrictions. *Transmissivity* is thereby established as a reliable, multiscale design parameter for engineering transport properties - analogous to nature's use of hierarchical structures to achieve optimized functionality (Fig. 1d).

The derived analytical framework quantitatively interprets simulation results and shows explicitly that the effect of geometric modulation on temperature difference *ΔT*, electrical and thermal resistances, TE efficiency, and output power is governed by a single function, *g(Tr)*. This function, defined as the ratio $G_{th}/G_{th}^0$, represents the conductance of a constricted geometry relative to a uniform-width counterpart. Crucially, this relationship is independent of carrier type, material choice, exact *constricted* geometry profile and operating conditions.

The universal scaling formalism provides a rigorous framework for evaluating TE performance in *constricted* geometries and establishes the foundation for a unified optimization strategy for composite TE legs that incorporate both constricted materials and contact.

Although *constricted* geometries consistently enhance TE efficiency compared to uniform-width structures - primarily through increased *ΔT* under convective operating conditions - this enhancement does not always translate into higher output power. The analytical scaling formalism for the output power density together with its graphical representation, constitute *TE Performance Design Maps* - practical, predictive tools that identify the conditions under which *constricted* geometries can enhance TE performance. Furthermore, an analytical optimization criterion is established: maximum TE performance is achieved at an optimal *Transmissivity, $Tr_{opt}$*, where the function *g($Tr_{opt}$)* equals the Biot number (*hL/k*).

This work provides the theoretical framework for multiscale design and optimization of *constricted* geometries, thereby enabling systematic exploration of design strategies for next-generation TE modules based on advanced thermoelectric metamaterials.

**APPENDIX A:**

**Analytical formalism for the temperature difference *ΔT* in uniform and non-uniform materials under convective operation conditions.**

Let us consider a uniform cuboid material of cross-sectional area *A* and length *L*. For a hot-side temperature $T_h$, the cold-side temperature $T_c$ under convective conditions is determined by the solution of the 1D steady-state thermal conduction equation:

$$\frac{d^2T}{dx^2} = 0 \Rightarrow T(x) = C_1 T + C_2 \tag{A.1}$$

where constants $C_1$ and $C_2$ are determined applying the boundary conditions:

At x=0: $\quad T(0) = T_h$

At x=L: $\quad -k\frac{dT}{dx}\Big|_{x=L} = h(T(L) - T_a)$

where $k$ is the material thermal conductivity, $h$ is the convection coefficient and $T_a$ is the ambient temperature. This gives:

$$T_c = T_h - \frac{hL}{k + hL}(T_h - T_a) \tag{A.2}$$

For corresponding non-uniform materials, Eq. (A.2) can be extended as follows:

$$T_c = T_h - \frac{hL}{k_{eff} + hL}(T_h - T_a) \tag{A.3}$$

in terms of the effective thermal conductivity $k_{eff}$, defined from the non-uniform material's thermal conductance $G_{th}$ through the relation:

$$G_{th} = k_{eff}\frac{A}{L} \tag{A.4}$$

The intrinsic thermal conductivity $k$ and the thermal conductance of the uniform cuboid material $G_{th}^0$ are related by the expression:

$$G_{th}^0 = k\frac{A}{L} \tag{A.5}$$

Combining Eqs (A.4) and (A.5) yields:

$$\frac{k_{eff}}{k} = \frac{G_{th}}{G_{th}^0} = \frac{R_{th}^0}{R_{th}} \tag{A.6}$$

Then, from Eqs. (A.3) and (A.6), it is obtained:

$$\Delta T = \frac{hL/k}{k_{eff}/k + hL/k}(T_h - T_a) \tag{A.7}$$

or equivalently expressed, in terms of $\Delta T_{max} \equiv T_h - T_a$:

$$\Delta T = \frac{hL/k}{G_{th}/G_{th}^o + hL/k}\Delta T_{max} = \frac{hL/k}{R_{th}^0/R_{th} + hL/k}\Delta T_{max} \tag{A.8}$$

## Data Availability Statement

The data that support the findings of this study are included in the article. Additional data are available by the author on request.

## Conflict of Interest

The author declares no competing Interests

## Funding

Research received no additional funding